\documentclass[twocolumn,english,aps,PRL,reprint, superscriptaddress,showpacs,longbibliography,showkeys]{revtex4-2}
\usepackage{amsmath,amssymb,bbm,mathrsfs,bm,braket,color,graphicx,comment}
\usepackage[colorlinks,citecolor=blue,urlcolor=blue,linkcolor = blue]{hyperref}
\usepackage[mathscr]{euscript}

\usepackage{hyperref}

\newcommand\minus{
  \setbox0=\hbox{-}
  \vcenter{
    \hrule width\wd0 height \the\fontdimen8\textfont3
  }%
}

\def\upket{\,\mid\uparrow\rangle}
\def\downket{\,\mid\downarrow\rangle}

\makeatother

\usepackage{tikz}

\newcommand{\squaredots}{
    \vspace{-.175em}
    \tikz[line cap=round, line join=round]{
    \draw[black] (0ex,0ex) -- (0ex,0.8ex) --  (0.8ex,0.8ex) --  (0.8ex,0ex) -- cycle;
    \draw[color=black, fill=black] (0ex,0ex) circle (0.175ex);
    \draw[color=black, fill=black] (0ex,0.8ex) circle (0.175ex);
    \draw[color=black, fill=black] (0.8ex,0ex) circle (0.175ex);
    \draw[color=black, fill=black] (0.8ex,0.8ex) circle (0.175ex);
    }
}

\newcommand{\squaredotsempty}{
    \tikz[line cap=round, line join=round, scale = 1.8, baseline=0pt]{
    \draw[black] (0ex,0ex) -- (0ex,0.8ex) --  (0.8ex,0.8ex) --  (0.8ex,0ex) -- cycle;
    \draw[color=black, fill=white] (0ex,0ex) circle (0.175ex);
    \draw[color=black, fill=white] (0ex,0.8ex) circle (0.175ex);
    \draw[color=black, fill=white] (0.8ex,0ex) circle (0.175ex);
    \draw[color=black, fill=white] (0.8ex,0.8ex) circle (0.175ex);
    }
}

\newcommand{\squaredotsonefilled}{
    \tikz[line cap=round, line join=round, scale = 1.8, baseline=0pt]{
    \draw[black] (0ex,0ex) -- (0ex,0.8ex) --  (0.8ex,0.8ex) --  (0.8ex,0ex) -- cycle;
    \draw[color=black, fill=white] (0ex,0ex) circle (0.175ex);
    \draw[color=black, fill=black] (0ex,0.8ex) circle (0.175ex);
    \draw[color=black, fill=white] (0.8ex,0ex) circle (0.175ex);
    \draw[color=black, fill=white] (0.8ex,0.8ex) circle (0.175ex);
    }
}

\begin{document}

\title{Quantum optimization via four-body Rydberg gates}

\author{Clemens Dlaska}
\affiliation{Institute for Theoretical Physics, University of Innsbruck, A-6020 Innsbruck, Austria}
\affiliation{Institute for Quantum Optics and Quantum Information of the Austrian Academy of Sciences, A-6020 Innsbruck, Austria}

\author{Kilian Ender}
\affiliation{Institute for Theoretical Physics, University of Innsbruck, A-6020 Innsbruck, Austria}
\affiliation{Parity Quantum Computing GmbH, A-6020 Innsbruck, Austria}

\author{Glen Bigan Mbeng}
\affiliation{Institute for Theoretical Physics, University of Innsbruck, A-6020 Innsbruck, Austria}

\author{Andreas~Kruckenhauser}
\affiliation{Institute for Quantum Optics and Quantum Information of the Austrian Academy of Sciences, A-6020 Innsbruck, Austria}
\affiliation{Center for Quantum Physics, Faculty of Mathematics, Computer Science and Physics, University of Innsbruck, 6020 Innsbruck, Austria}

\author{Wolfgang Lechner}
\affiliation{Institute for Theoretical Physics, University of Innsbruck, A-6020 Innsbruck, Austria}
\affiliation{Parity Quantum Computing GmbH, A-6020 Innsbruck, Austria}

\author{Rick van Bijnen}
\affiliation{Institute for Quantum Optics and Quantum Information of the Austrian Academy of Sciences, A-6020 Innsbruck, Austria}
\affiliation{Center for Quantum Physics, Faculty of Mathematics, Computer Science and Physics, University of Innsbruck, 6020 Innsbruck, Austria}

\begin{abstract} 
There is a large ongoing research effort towards obtaining a quantum advantage in the solution of combinatorial optimization problems on near-term quantum devices. A particularly promising platform for testing and developing quantum optimization algorithms are arrays of trapped neutral atoms, laser-coupled to highly excited Rydberg states.  However, encoding combinatorial optimization problems in atomic arrays is challenging due to the limited inter-qubit connectivity given by their native finite-range interactions. Here we propose and analyze a fast, high fidelity four-body Rydberg parity gate, enabling a direct and straightforward implementation of the Lechner-Hauke-Zoller (LHZ) scheme and its recent generalization, the parity architecture, a scalable architecture for encoding arbitrarily connected interaction graphs. Our gate relies on onetime-optimized adiabatic laser pulses and is fully programmable by adjusting two hold-times during operation. We numerically demonstrate an implementation of the quantum approximate optimization algorithm (QAOA) for a small scale test problem. Our approach allows for efficient execution of variational optimization steps with a constant number of system manipulations, independent of the system size, thus paving the way for experimental investigations of QAOA beyond the reach of numerical simulations.

\end{abstract}
\maketitle

Currently available quantum devices are capable of generating controlled dynamics beyond the reach of numerical simulations on even the most powerful classical supercomputers \cite{Arute2019, Zhong2020}. These quantum devices will have up to a few hundred qubits available, without error correction, and have been termed Noisy Intermediate Scale Quantum (NISQ) devices. A key challenge for the field of quantum technology at this very moment is to find ways of putting the computational power of near-term quantum devices to good use \cite{Preskill2018, Deutsch2020}. 
In this era of NISQ devices, the development of specialized algorithms, targeting specific problems that provide a structural match with the strengths of a particular quantum platform, is thus highly desirable. A strategy of co-design of algorithms and experimental platforms aims at developing scientifically and industrially relevant applications in the near term, before the need for error correction arises. 

Here we focus on designing specialized quantum hardware for solving combinatorial optimization problems, using neutral atoms trapped in tweezer arrays, laser-coupled to highly excited Rydberg states~\cite{Bernien2017, Barredo2018,Pichler2018, Levine2019, Graham2019, Madjarov2020,Scholl2020, Ebadi2020, Semeghini2021, Bluvstein2021}. The Rydberg states provide strong and tunable interactions, that can be switched on and off by coherently coupling ground and Rydberg states.
Combined with single particle operations, the interactions form appealing building blocks for the QAOA~\cite{Farhi2014, Farhi2016}. There, the goal is to find approximate solutions to combinatorial optimization problems, cast in the form of energy minimization of a general $N$-spin problem Hamiltonian
\begin{equation}
    \hat{H}_P = \sum_{i<j}^NJ_{ij} \hat{\sigma}_z^{(i)} \hat{\sigma}_z^{(j)} + \sum_{i<j<k}^NJ_{ijk} \hat{\sigma}_z^{(i)} \hat{\sigma}_z^{(j)}\hat{\sigma}_z^{(k)} + \dots, 
\end{equation}
where $\hat{\sigma}_{\lbrace x,y,z\rbrace}$ denote the Pauli spin operators and $\lbrace J_{ij}, J_{ijk},\dots \rbrace$ are infinite-range interactions. The QAOA attempts to find low energy solutions, by driving a system of quantum spins alternatingly with a driver Hamiltonian $\hat{H}_X = \sum_i \hat{\sigma}_x^{(i)}$ and the problem Hamiltonian $\hat{H}_P$. Due to the heuristic nature of this algorithm its practical performance in a regime beyond the capability of classical computers is difficult to predict and requires to be experimentally tested \cite{Zhou2020}. 
Recent advances in Rydberg experiments, such as coherent control of atomic states and deterministic atom positioning of hundreds of atoms, make the Rydberg platform a particularly promising target for such investigations. 

Direct experimental implementations of QAOA with Rydberg atoms are, however, limited by the binary nature of the Rydberg interaction and their polynomially decaying interaction strengths, which only admit scalable experimental implementations of $\hat{H}_P$ for very specific problems~\cite{Pichler2018, Song2021}. 

\begin{figure}[t]
\begin{centering}
\includegraphics[width=\columnwidth]{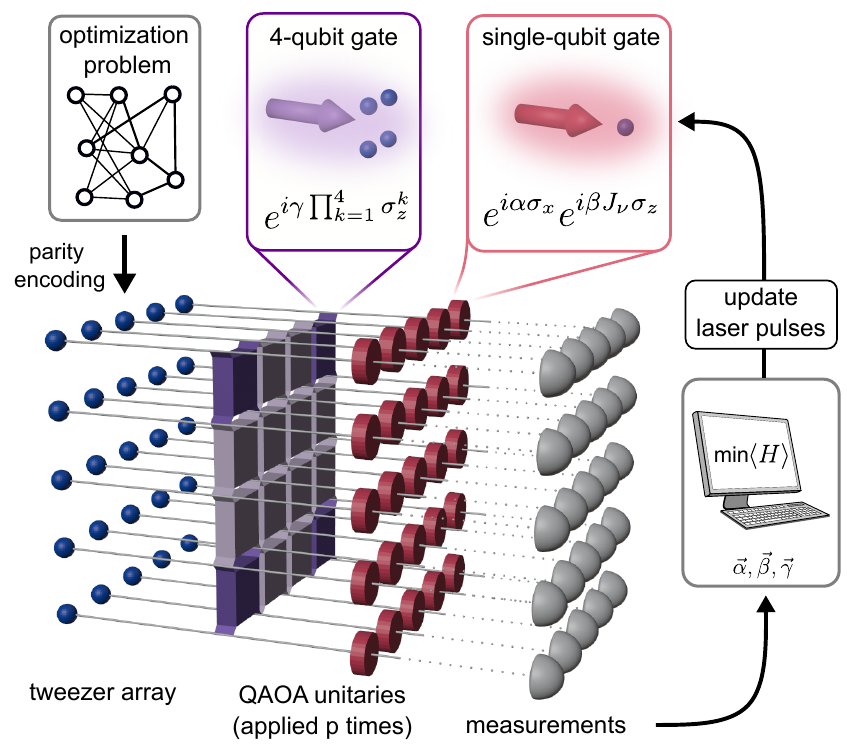}
\par\end{centering}
 \protect\caption{\textit{Rydberg parity QAOA protocol}. Arbitrarily connected optimization problems can be parity encoded in a regular geometry of neutral atoms trapped in e.g.\@ optical tweezers. After initializing the Rydberg quantum processor in an equal superposition state, generating variational wave functions by applying QAOA unitaries only requires local control of laser fields generating quasi-local four- (square boxes) and single-qubit gates (discs).}
\label{fig1}
\end{figure}

Instead of attempting to directly engineer the spin model version of $\hat{H}_P$, we adopt the parity architecture~\cite{Lechner2015,Ender2021}, a scalable and problem independent quantum hardware blueprint for generic combinatorial optimization problems. 
Running the QAOA then only requires problem dependent single-qubit gates and problem independent multi-qubit phase-gates acting on three or four qubits at the corners of  $2\times2$ plaquettes as (see Fig.~\ref{fig1}) 
\begin{equation}\label{eq:gate}
    U_{\squaredots}(\gamma) = e^{i\gamma\prod_{k}\hat{\sigma}_z^{(k)}},
\end{equation}
where the latter do not naturally exist on the Rydberg platform.

In the following we show how such a gate can be directly engineered between ground state atoms utilizing time-optimal adiabatic laser-coupling to Rydberg states, i.e. without relying on (distinct species) auxiliary qubits~\cite{Glaetzle2017} or decomposition into two-body gates~\cite{Lechner2020}. We provide a simple two-pause strategy to program arbitrary phases $\gamma$ subsequent to a onetime optimization of laser-ramps within parameter-limits given by particular experiments. The entire QAOA algorithm can then be implemented on present-day experiments as an optimization of the duration of direct laser pulses. Below we explain the details and performance of our scheme, and give a numerical demonstration of the QAOA protocol on the Rydberg platform.

\begin{figure*}[t]
\begin{centering}
\includegraphics[width=\textwidth]{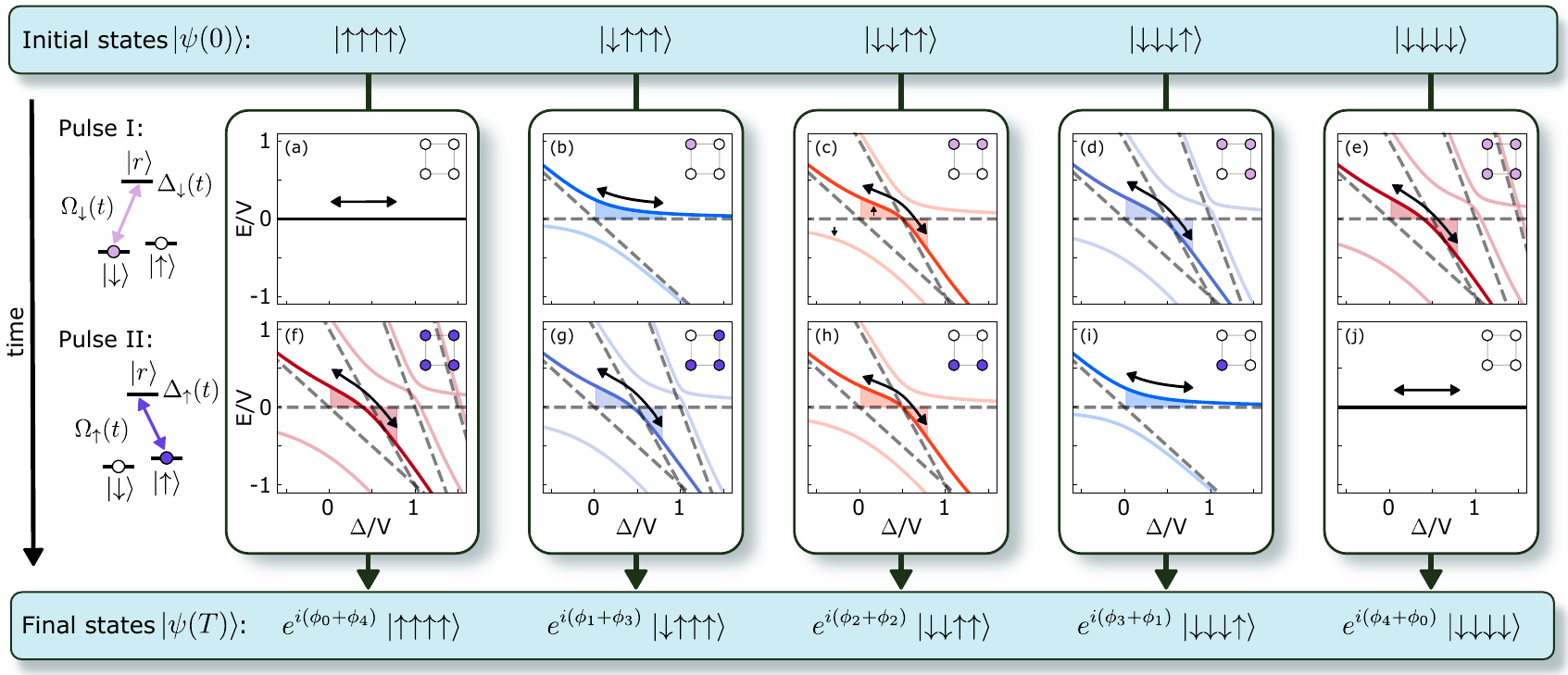}
\par\end{centering}
\protect\caption{Four-body Rydberg gate protocol. The laser-parameter dependent plaquette energy-spectrum exhibits distinct behavior w.r.t.\@ the number of laser-excitable spins (indicated by\protect\squaredotsempty,\protect\squaredotsonefilled, \dots). Solid (dashed) lines show the energy-spectrum as function of the laser detuning $\Delta_{\downarrow,\uparrow}$ for a fixed Rabi-frequency of  $\Omega_{\downarrow,\uparrow} = V/2$ ($\Omega_{\downarrow,\uparrow} = 0$). Applying an adiabatic, time dependent laser pulse $(\Omega_{\downarrow}(t),\Delta_{\downarrow}(t))$ addressing qubit state $\ket{\downarrow}$ imprints an excitation-sector dependent dynamical phase $\phi_n$ on the corresponding computational basis states (upper row). For constant Rydberg interaction strengths between plaquette atoms, subsequently applying the same adiabatic laser pulse on qubit states $\ket{\uparrow}$ leads to the desired phase separation of even and odd parity configurations (bottom line).}
\label{fig2new}
\end{figure*}


\subsection*{Rydberg parity QAOA}
The parity hardware architecture provides a blueprint for a problem independent and scalable quantum processor that is tailored to tackle generic combinatorial optimization problems (see Appendix~\ref{appendix:encoding} for a detailed introduction). In short, parity-qubits encode the relative orientation, i.e.\@ the parity, of spins representing the optimization problem, with $J_{ij}\hat{\sigma}_z^{(i)}\hat{\sigma}_z^{(j)}\rightarrow J_\mu \hat{\sigma}_z^{(\mu)}$, $J_{ijk}\hat{\sigma}_z^{(i)}\hat{\sigma}_z^{(j)}\hat{\sigma}_z^{(k)}\rightarrow J_\nu \hat{\sigma}_z^{(\nu)}$ etc., replacing infinite-range interactions $\lbrace J_{ij}, J_{ijk},\dots\rbrace $ by local-fields $\lbrace J_\mu, J_\nu,\dots \rbrace$. Since the parity transformation increases the number of qubits to the number $K$ of interactions present in the optimization problem, the original $N$-qubit code-space needs to be stabilized by quasi-local three- or four-qubit stabilizers of the form ${H_{\squaredots} \propto \prod_{k=1}^l\hat{\sigma}_z^{(k)}}$ (i.e.\@ $l=3,4$), that act as energetic constraints on $2\times2$ plaquettes \cite{Rocchetto2016, Ender2021}. 

Implementations of QAOA for parity encoded optimization problems rely on alternatingly driving the quantum spin system, prepared in the $\ket{+}^{\otimes K}$ state, with a driving Hamiltonian and the problem Hamiltonian. While the single qubit driving Hamiltonian $\hat{H}_X = \sum_\nu^K \hat{\sigma}_x^{(\nu)}$ remains as before, the problem Hamiltonian $\hat{H}_P$ is now decomposed into a single qubit problem encoding $\hat{H}_Z = \sum_\nu^K J_{\nu} \hat{\sigma}_z^{(\nu)}$, and  a quasi local constraint term $\hat{H}_C=\sum_{\squaredots} H_{\squaredots}$, where the sum runs over all $2\times2$ plaquettes denoted by $\squaredots$. Alternatingly applying each of the Hamiltonian operators $p$ times, the QAOA thus generates states
\begin{equation}\label{eq:QAOA}
\ket{\psi} = \prod_{j=1}^p e^{-i\alpha_j \hat{H}_X}e^{-i\beta_j \hat{H}_Z}e^{-i\gamma_j \hat{H}_C}\ket{+}^{\otimes K},
\end{equation}
where variational parameters $\alpha_j, \beta_j$, and $\gamma_j$ $(j = 1,2,\dots,p)$ determine the duration of driving with $\hat{H}_X, \hat{H}_Z, \hat{H}_C$, respectively. Low energy solutions to the original optimization problem are then determined in a quantum-classical feedback loop, where a classical computer optimizes the parameters $(\bm{\alpha}, \bm{\beta}, \bm{\gamma})$, based on measurements of the energy $\bra{\psi}(\hat{H}_Z + \hat{H}_C)\ket{\psi}$ in the state $\ket{\psi} = \ket{\psi(\bm{\alpha}, \bm{\beta}, \bm{\gamma})}$.

The quantum spin system we have in mind consists of a regular array of trapped neutral atoms, e.g.\@ Rubidium (${}^{87}\mathrm{Rb}$) atoms trapped in optical tweezers (see Fig.~\ref{fig1}). Each atom realizes a qubit by encoding the qubit basis $\lbrace\downket{},\upket{}\rbrace$ in a pair of atomic ground states (e.g.\@ two hyperfine states). 
We assume the ability to locally address atoms with targeted laser light, e.g.\@ by using spatial light modulators (SLMs) \cite{Fukuhara2013, Bijnen2015, Labuhn2016, Ebadi2020}. The single particle operations $\hat{H}_Z$ can then be implemented through AC Stark shifts from laser coupling to low-lying excited states. The driver Hamiltonian $\hat{H}_X$ can be implemented through Raman transitions. In the following, we discuss the Rydberg implementation of the key nontrivial component, the many-body phase gate $e^{-i\gamma_j \hat{H}_C}$.

\subsection*{Four-qubit parity gate implementation}
The main challenge for experimental realizations of the parity-QAOA algorithm is a direct and straightforward implementation of the four-qubit gate $U_{\squaredots}(\gamma) = e^{-i \gamma H_{\squaredots}}$. The operator $H_{\squaredots}$ energetically separates plaquette states $\ket{z_\mathrm{even}}$ with an even number of particles in the $\ket{\downarrow}$ state, from plaquette states $\ket{z_\mathrm{odd}}$ with an odd number of particles in the state $\ket{\downarrow}$.
The desired gate operation $U_{\squaredots}(\gamma)$ thus corresponds to a four-qubit phase gate, mapping the plaquette states as follows:
\begin{equation}\label{eq:gate_detail}
\begin{split}
    U_{\squaredots}(\gamma)\ket{z_\mathrm{odd}}&= e^{i\gamma} \ket{z_\mathrm{odd}},\\
    U_{\squaredots}(\gamma)\ket{z_\mathrm{even}}&= e^{-i\gamma}\ket{z_\mathrm{even}}.
\end{split}
\end{equation}
We will show that this operation can be implemented with two adiabatic laser pulses, with time-dependent intensity and detuning, where a first pulse couples only the $\ket{\downarrow}$ states to a Rydberg level $\ket{r}$, and the second pulse couples only the $\ket{\uparrow}$ states to the same Rydberg state $\ket{r}$ (see Fig.~\ref{fig2new}). The first pulse gives all plaquettes with $n$ particles in $\ket{\downarrow}$ a phase $\phi_n$, whereas the second pulse gives all plaquettes with $4-n$ particles in the $\ket{\uparrow}$ state a phase $\phi_{4-n}$. Due to the Rydberg-Rydberg interaction between atoms in state $\ket{r}$, the phases can be programmed to satisfy $\phi_1 + \phi_3 = \gamma$, and $\phi_0 + \phi_4 = 2\phi_2 = -\gamma$.

We assume that the $2\times2$ plaquettes can be individually addressed with a Rydberg excitation laser with a time-dependent Rabi frequency $\Omega(t)$, and detuning $\Delta(t)$, e.g.\@ using SLMs. The first pulse couples only the $\ket{\downarrow}$ to the Rydberg state, and in this case the relevant Hamiltonian is of the form 
\begin{equation}\label{eq:H2x2}
\begin{split}
    \hat{H}_{2\times2} &= \sum_i \left[ -\Delta(t) \ket{r_i}\bra{r_i} + \frac{\Omega(t)}{2} \ket{r_i}\bra{\downarrow} + \mathrm{H.c.}\right]\\ &+ \sum_{i<j} V_{ij} |r_ir_j \rangle\langle r_ir_j|,
\end{split}
\end{equation}
where the sums run over the indices $i, j$ on the plaquette, and $V_{ij}$ is the van der Waals interaction energy between atoms $i$ and $j$. Since the desired gate operation is permutation symmetric, it is beneficial to have $V_{ij} = V$. We therefore assume here that the plaquette atoms form a tetrahedral configuration, which can be realised by displacing every second lattice diagonal out of a square lattice. We note however, that this is not a strict requirement~\footnote{For atoms placed in a square lattice geometry the scenario for two excitation particles will split up into two cases due to the difference in interaction strengths between atoms in the Rydberg state on the sides vs. the diagonals of the plaquette. That means, that in comparison to the tetrahedral configuration an additional phase needs to be controlled, which does not pose a limitation on the scheme}.

For designing our phase gate, we will exploit properties of the many-body eigenspectrum of Eq.~(\ref{eq:H2x2}).
Since (during the first pulse) we are only coupling the $\ket{\downarrow}$ states, the particles in $\ket{\uparrow}$ remain in a noninteracting ground state, and do not participate in the dynamics. In particular, they can not be excited to the Rydberg state, and the number of relevant eigenstates of Eq.~(\ref{eq:H2x2}) is therefore dependent on the number of particles originally in state $\ket{\downarrow}$ at the start of the pulse, as illustrated in Fig.~\ref{fig2new}, top row. Dashed lines indicate product eigenstates in the limit $\Omega = 0$, whereas solid, colored, lines are the eigenstates for a specific $\Omega > 0$.
If there are initially no particles in $\ket{\downarrow}$ [panel (a)], there is just one eigenstate, i.e. $\ket{\uparrow\uparrow\uparrow\uparrow}$. 
If there is one particle in the plaquette in state $\ket{\downarrow}$, an additional eigenstate appears, with one particle in $\ket{r}$. For $\Omega = 0$ (dashed line), this state has an energy $-\Delta$. For $\Omega>0$ this state forms an anticrossing with the original product ground state. Similarly, for $n = 2, 3, 4$ particles in $\ket{\downarrow}$ [panels (c), (d), (e)], more coupled eigenstates with $n$ particles in $\ket{r}$ appear with slopes $-n\Delta$. Moreover, these states have an energy offset at $\Delta = 0$ due to the interactions, equal to $n (n - 1)V / 2$. It should be noted that the widths of the anticrossings in the spectrum also increase with $n$.

We now design pulses $\{ \Omega(t), \Delta(t) \}$, adiabatically connecting the initial product state of ground state atoms to one of the many-body eigenstates. The initial value of the detuning at $t=0$ and $\Omega(0) = 0$ determines to which of the eigenstates we connect when increasing $\Omega > 0$. For example, $\Delta(0) < 0$ connects us to the lowest eigenstate, and $0< \Delta(0) < V/2$ connects to the first excited state [small arrows in Fig.~\ref{fig2new}(c)]. For illustrative purposes, we operate on the first excited many-body state. After adiabatically increasing $\Omega > 0$ and subsequently sweeping the detuning back and forth (arrows in Fig.~\ref{fig2new}), the plaquettes pick up a dynamical phase (the time integral of the particular eigenenergy-trajectory, indicated as shaded areas in Fig.~\ref{fig2new}), which is dependent on the number $n$, due to the many-body spectrum depending on $n$. Since the desired gate operation is invariant under global spin-flips, we can achieve the desired effect that the odd and even plaquettes pick up equal and opposite phases $\pm\gamma$ by repeating the pulse in exactly the same fashion, but this time coupling the $\ket{\uparrow}$ ground states to the Rydberg state (panels (f) - (j) in Fig.~\ref{fig2new}). 
By simultaneously illuminating plaquettes that are separated by a line of non-illuminated atoms (see highlighted plaquettes in Fig.~\ref{fig1}), i.e.\@ to avoid crosstalk between plaquettes, the whole many-body phase gate $e^{-i\gamma_j \hat{H}_C}$ can be realized in 9 illumination rounds independent of the system size. In the following section we discuss a strategy to efficiently control the precise values of $\gamma$ that is particularly suitable for usage in variational quantum algorithms.

\subsection*{Two pause protocol}

Ideal co-design of algorithms and hardware requires a close connection between algorithmic variables and available hardware manipulations, i.e.\@ minimal computational and experimental overhead to realize these variables in hardware. For the particular case of our four-body gate implementation this requirement can be fulfilled by a straight forward two-pause protocol that only needs onetime optimization and calibration of laser pulse-shapes. 

\begin{figure}[th]
\begin{centering}
\includegraphics[width=\columnwidth]{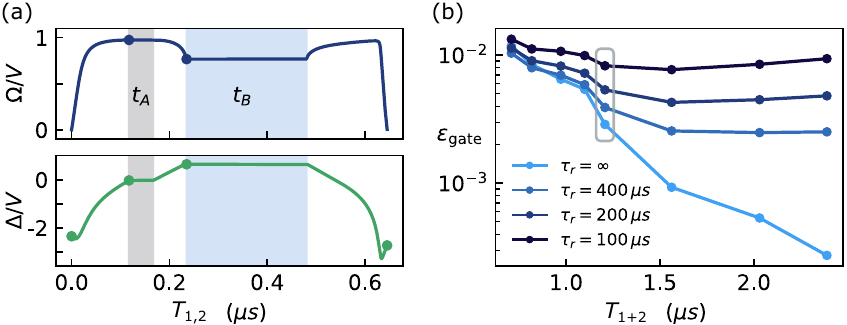}
\par\end{centering}
\protect\caption{(a) Two pause pulse example. Laser parameters that are used for pulse optimization are indicated as dots (see main text). (b) Gate error of the parity gate for experimental conditions as: $V = \Omega_\mathrm{max} =  2\pi\times 40\,\mathrm{MHz}$,  $\Delta_{\text{start},\text{end}}/V \in \left[-3.0,0.0\right[$,  $\Delta_{A,B}/V \in \left[-3.0,1.0\right[$, averaged over $10^4$ randomly chosen phase-combinations (see Appendix~\ref{appendix:TPP}). The highlighted points correspond to the pulse shown in panel (a) where the pause-times $t_{A,B}$ are adjusted such that $\gamma = \pi$.}
\label{fig3}
\end{figure}

Our protocol relies on adiabatic trajectories $(0,  \Delta_\text{start})\rightarrow (\Omega_A, \Delta_A) \rightarrow(\Omega_B, \Delta_B)\rightarrow(0, \Delta_\text{end})$, where the corresponding laser-pulse is held (``paused") at $(\Omega_{A,B}, \Delta_{A,B})$ for durations $t_{A,B}$ [see\@ Fig.~\ref{fig3}(a)]. The key observation is that for an arbitrary gate phase $\gamma$ in Eq.~(\ref{eq:gate_detail}), there exists an analytic solution of hold times $t_{A,B}$, realizing the desired phase (see Appendix~\ref{appendix:TPP}). The precise solutions, and hence the total gate duration, depend on the values of $\mathbf{\Omega} = (\Omega_A,\Omega_B)$ and $\mathbf{\Delta}= (\Delta_\text{start},\Delta_A, \Delta_B,\Delta_\text{end})$, and the adiabatic path connecting them.

We determine the waypoints $(\mathbf{\Omega}, \mathbf{\Delta})$ of the adiabatic path by numerically optimizing the total gate duration for all values of $\gamma \in [0, 2\pi]$, for the worst case scenario, and given experimental constraints such as achievable interaction strengths $V$ and maximum Rabi frequencies $\Omega$. The paths $\Omega(t), \Delta(t)$, connecting the waypoints $(\mathbf{\Omega}, \mathbf{\Delta})$, are calculated using a novel numerical approach based on quantum adiabatic brachistochrones (QAB)~\cite{Roland2002,Jansen2007,Rezakhani2009} (see Appendix~\ref{appendix:QAB}). Once this one-time optimization is done, executing the QAOA consists of only varying the hold times $t_{A, B}$, irrespective of the precise problem.

\subsection*{Gate performance}

In this section we assess the performance of our parity gate protocol for a realistic experimental scenario.
We assume an interaction strength $V = 2\pi \times 40 \mathrm{MHz}$, e.g. achievable for $68S$ states of ${}^{87}\mathrm{Rb}$,  and particles spaced at $5 \mathrm{\mu m}$. In  Appendix~\ref{appendix:Rydberg}  we provide a detailed discussion of potential considerations, including three- and four-body effects. We note that for interaction strengths of this magnitude and $\mu s$ gate operation times, trapping with about $1-2 \mathrm{mK}$ deep traps of the Rydberg states would be required \cite{Mukherjee2011, Barredo2020trapping}. The lifetime of the $68S$ states at 300~K is about $150 \mathrm{\mu s}$ \cite{BeterovLifetimes}.

For parameters in this regime, Fig.~\ref{fig3}(b) analyzes the average gate-error $\epsilon_\mathrm{gate}=1-\overline{\mathcal{F}}$, where $\overline{\mathcal{F}}$ denotes the mean of the average gate fidelity over $10^4$ gate realizations, i.e.~$\gamma$-values. We optimized the laser-parameters in the coherent, i.e.\@ noiseless, regime for various levels of adiabaticity using a 100-steps basin-hopping algorithm \cite{Wales1997}. There, the gate error [see light blue line in Fig.~\ref{fig3}(b)] solely originates from diabatic errors and thus can be arbitrarily reduced by making the gate more adiabatic, i.e.\@ slower. However, the finite lifetime of Rydberg states restricts the maximal gate-duration and thus limits the achievable gate-performance. Including dissipation (see Appendix~\ref{appendix:ME}) shows that the best possible gate-performance is a trade-off between diabatic and dissipative error mechanisms [see Fig.~\ref{fig3}(b)].

\subsection*{QAOA simulations}
\begin{figure}[t]
\begin{centering}
\includegraphics[width=\columnwidth]{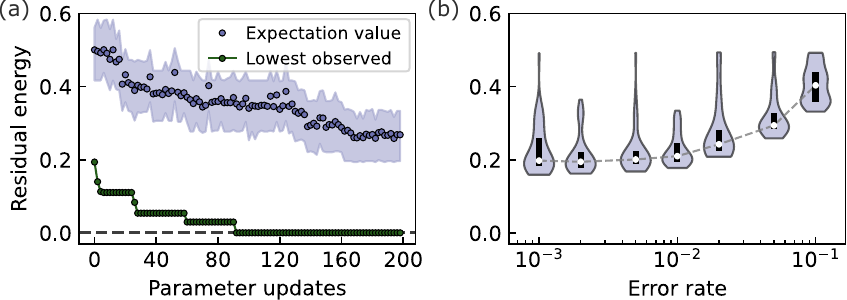}
\par\end{centering}
\protect\caption{
(a) Example QAOA simulation for 20 qubits. Shown are the lowest observed- and average residual energy [cf.\@ Eq.~\ref{eqn:res_en}] after each parameter update (i.e.\@ before acceptance/rejection).
(b) Distribution of average residual energies of 50 independent optimization runs for a single optimization problem with varying error rates of the four-body parity gate. The black bars visualize the 25th to 75th percentiles and the white circles denote the median of the distribution.}
\label{fig4a}
\end{figure}

We numerically demonstrate the feasibility of our parity-QAOA implementation on small test-scale problems of $K=20$ qubits (see Fig.~\ref{fig4a}). In particular we focus on a qubit array arranged in a $4 \times 5$ grid where local fields $J_j$ are randomly chosen to be either $-1$ or $1$. This corresponds to a small example logical optimization problem of a bipartite graph involving 9 logical qubits. The main objective of our simulation is to investigate the robustness of our QAOA scheme under varying (depolarizing) noise levels of the four-qubit parity gate. To this end we numerically simulated the QAOA circuit Eq.~\eqref{eq:QAOA} with circuit depth $p=3$ for various error rates of the four-body gate, while keeping the single-qubit error rates constant at $0.05\%$. More precisely, after initialization in an equal superposition of all computational basis states, the parameters are stochastically updated in a quantum-classical feedback loop. Starting with all parameters $\{\alpha, \beta, \gamma \}$ being zero, a randomly chosen parameter is updated at each optimization step which consists of 500 circuit executions and subsequent measurements in the $z$-basis in order to estimate the energy of the system. The parameter update is accepted if the estimated energy expectation value $\langle E \rangle$ is decreased, otherwise rejected. Figure~\ref{fig4a}(a) shows the residual energy
\begin{equation}
    E_\text{res} = \frac{\langle E \rangle-E_\text{min}}{E_\text{max}-E_\text{min}}
    \label{eqn:res_en}
\end{equation} 
as function of the number of parameter updates for a sample experiment with a four-qubit error rate of $0.01\%$. 

Figure~\ref{fig4a}(b) shows the distribution of the residual energies of 50 independent optimization runs for a single instance of randomly chosen couplings for varying noise levels. During optimization we allowed for 200 parameter updates, where $\langle E \rangle$ is estimated upon 500 circuit executions and energy measurements. For the final values shown in Fig.~\ref{fig4a}(b) each optimized circuit is repeated 5000 times for a better energy estimate. We observe that the performance is robust against error rates up to a few percent which can be achieved with sub-$\mu$s Rydberg gate protocols [see Fig.~\ref{fig3}(b)].

\subsection*{Conclusions and Outlook} 
Solving optimization problems using a variational gate model approach is of current interest among different qubit platforms and architectures. While present day Rydberg experiments have seen enormous progress in quantum state control and particle numbers, their focus has been so far predominantly on quantum simulation. Our proposed Rydberg parity gate will enable these experiments to explore solving arbitrary combinatorial optimization problems, providing a new direction towards quantum computing tasks, without requiring substantial hardware changes. 

The proposal builds on the parity mapping, which translates optimization problems with high connectivity,  higher-order interactions and constraints to a representation on a simple square lattice where, notably, all required interactions are independent of the problem and only act among nearest neighbors. This provides a natural fit with available Rydberg operations. In particular, we show that in our approach the variational parameters for QAOA are simply hold times of laser pulses, during which the quantum states accumulate phases. Furthermore, the inherent scalability of the parity architecture is naturally complemented by the scalability of the Rydberg platform. Increasing system size does not require modifications to the parity gate, nor fundamental modifications on the hardware side.  

We expect intermediate scale Rydberg experiments with up to hundreds of atoms to be able to investigate the performance of QAOA on the parity architecture, in regimes where its performance cannot be investigated by classical simulations anymore. Going beyond these system sizes, the Rydberg lifetimes will become an issue, and will require either much larger interaction strengths for faster gate operations, or modified implementations suited for circular Rydberg states with much longer lifetimes~\cite{BruneCircular, MeinertCircular, ThompsonCircular}.



\section*{Acknowledgments}
We thank A. M. Kaufman and H. Pichler for helpful discussions.
 The work is supported by the European Union program Horizon 2020 under Grants Agreement No.~817482 (PASQuanS), and by the Austrian Science Fund (FWF) through a START grant under Project No. Y1067-N27 and the SFB BeyondC Project No. F7108-N38, the Hauser-Raspe foundation. This material is based upon work supported by the Defense Advanced Research Projects Agency (DARPA) under Contract No. HR001120C0068. Any opinions, findings and conclusions or recommendations expressed in this material are those of the author(s) and do not necessarily reflect the views of DARPA.
\appendix
\section{Parity encoding of optimization problems}
\label{appendix:encoding}

By mapping the $k$-fold product of a subset of logical spins (denoted by $\hat{\sigma}$) onto a single physical parity qubit (denoted by $\tilde{\sigma}$), e.g.\@ $ J_{ijk}\, \hat{\sigma}^{(i)}_z \hat{\sigma}^{(j)}_z \hat{\sigma}^{(k)}_z \rightarrow J_\nu\, \tilde{\sigma}^{(\nu)}_z $, it is possible to directly encode combinatorial optimization problems with arbitrary higher-order $k$-body terms on a square lattice \cite{Ender2021}. 
In this mapping the optimization problem is fully controlled by the local fields $J_\nu$, while all interactions are geometrically local and problem independent. 

\begin{figure}[h]
\begin{centering}
\includegraphics[]{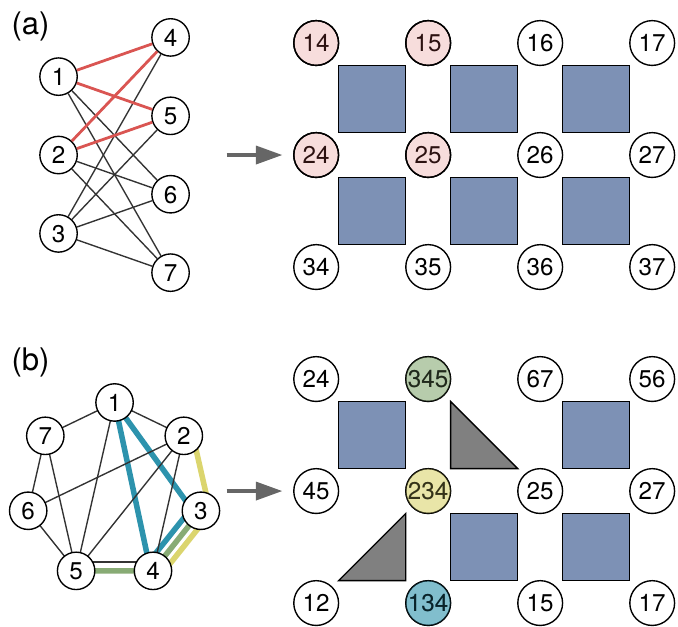}
\par\end{centering}
 \protect\caption{Examples of parity encoded optimization graphs.
 (a) Parity encoding of a logical bipartite graph (left panel) in a physical square lattice (right panel). Edges (black lines) between nodes (circles) in the logical graph denote two-body interactions. A closed loop in the logical graph and the corresponding plaquette of physical nodes are highlighted in red. Blue squares denote quasi-local four-body interactions. (b) Parity encoding of a logical hypergraph comprising two- and three-body interactions. The three-body interactions and their corresponding physical qubits are highlighted with colors. Grey triangles denote three-body interactions.}
\label{fig1_methods}
\end{figure}

An optimization problem in $N$ binary variables represented by a Hamiltonian with $K>N$ terms will be mapped onto $K$ such parity qubits. As the logical problem is now embedded in a higher-dimensional Hilbert space additional constraints are necessary to define the subspace containing the logical problem. If the optimization problem can be represented by a graph (i.e.\@ with only two-body interactions) these constraints can be constructed from closed loops therein [highlighted red in Fig.~\ref{fig1_methods}(a)]. Along such a closed loop in the logical graph only an even amount of parity changes can occur. In the parity encoding this is ensured by introducing an energy penalty of the form $H_{\squaredots} \propto - \tilde{\sigma}_z \tilde{\sigma}_z \tilde{\sigma}_z \tilde{\sigma}_z$ on a plaquette, energetically penalizing configurations which have an odd number of parity changes and thus no correspondence to the logical problem.

Another way of defining the constraints is to look for products of physical qubits such as $\tilde{\sigma}^{(14)}_{z}\tilde{\sigma}^{(15)}_{z}\tilde{\sigma}^{(25)}_{z}\tilde{\sigma}^{(24)}_{z}$, whose corresponding product in the logical spins has to be one, i.e.\@ ${\hat{\sigma}^{(1)}_{z}\hat{\sigma}^{(4)}_{z}\hat{\sigma}^{(1)}_{z}\hat{\sigma}^{(5)}_{z}\hat{\sigma}^{(2)}_{z}\hat{\sigma}^{(5)}_{z}\hat{\sigma}^{(2)}_{z}\hat{\sigma}^{(4)}_{z}=1}$. The energy penalties $H_{\squaredots}$ ensure that these products of physical qubits also fulfill this condition.
Notice that the indices of logical qubits on a plaquette always appear an even amount of times. This generalized definition of a constraint can be readily adapted to encode hypergraphs, i.e.\@ to encode arbitrary optimization problems including higher-order $k$-body terms. An example of this is shown in Fig.~\ref{fig1_methods}(b).

\section{Two pause protocol}
\label{appendix:TPP}
In the following we describe in detail the two pause protocol, which allows to program all phase-differences $\delta\phi_{a,b} =\phi_\mathrm{even}^{a,b} -\phi_\mathrm{odd}$, with  $\phi_\mathrm{even}^a = \phi_0 + \phi_4$ and $\phi_\mathrm{even}^b = 2\phi_2$ and $\phi_\mathrm{odd} = \phi_1 + \phi_3$, by adjusting two pause-times. The gate described in Eq.~\ref{eq:gate_detail} is realized when $\delta\phi_{a,b} = -2\gamma$. The laser-pulse we have in mind consists of five parts (three ramps and two pause-periods): 
\begin{itemize}
    \item[(i)]  ramp $(0, \Delta_\text{start})\rightarrow (\Omega_A, \Delta_A)$
    \item[(ii)] hold at $(\Omega_A, \Delta_A)$ for time $t_A$
    \item[(iii)] ramp  $(\Omega_A, \Delta_A)\rightarrow (\Omega_B, \Delta_B)$
    \item[(iv)] hold at $(\Omega_B, \Delta_B)$ for time $t_B$
    \item[(v)] ramp  $(\Omega_B, \Delta_B)\rightarrow (0, \Delta_\text{end})$
\end{itemize}
Associating a time $T_\nu$ to every step gives a total gate duration $T_\mathrm{gate} = 2\sum_{\nu=1}^5T_\nu$. The phase differences $\delta\phi_{a,b}$ can then be written as sum over the corresponding phase differences acquired during a particular step of the protocol
\begin{equation}
    \delta\phi_j = \sum_{\nu=1}^5\delta\phi_{j}^{(\nu)},\quad j=a,b.
    \label{eq:5phases}  
\end{equation}
Assuming the laser parameters $\bm\Omega = (\Omega_A, \Omega_B)$, $\bm\Delta = (\Delta_\text{start}, \Delta_A, \Delta_B, \Delta_\text{end})$ to be fixed and thus the ramps and the corresponding energies to be known, the only variables in equation~\eqref{eq:5phases} are the phases corresponding to the hold periods. These can be easily computed as
\begin{equation}
    \delta\phi_j^{(2)} = t_A\delta E_j^A, \quad \delta\phi_j^{(4)} = t_B\delta E_j^B,
\end{equation}
with $\delta E_j(\Omega,\Delta) = E_\mathrm{odd}(\Omega,\Delta) - E_\mathrm{even}^{(j)}(\Omega,\Delta)$ denoting the energy differences of the instantaneous eigenstates and $\delta E_{j}^\mu \equiv \delta E_j(\Omega_\mu,\Delta_\mu)$. Thus, the hold times $t_{A,B}$ leading to a desired phase combination ($\delta\phi_a, \delta\phi_b$) can be easily found by solving a system of linear equations given by
\begin{eqnarray}\label{eq:holdtimesequation}
\begin{split}
   \delta E_{a}^At_A+\delta E_{a}^Bt_B &= \delta\phi_a-\delta\phi_a^\mathrm{ramps},\\
     \delta E_{b}^At_A+\delta E_{b}^Bt_B &= \delta\phi_b-\delta\phi_b^\mathrm{ramps}, 
\end{split}
\end{eqnarray}
with $\delta\phi_j^\mathrm{ramps} =\sum_{\nu = 1,3,5} \delta\phi_j^{(\nu)}$. 
We recall that, the target phases $\delta\phi_j$, are only defined up to integer multiple of $2\pi$. We use this additional degree of freedom to replace $\delta\phi_j \to \delta\phi_j + 2\pi m_j$, with $m_j\in\mathbb{Z}$, in Eq.~\eqref{eq:holdtimesequation}. We thus generate a set of solutions $t_{A,B}(\delta\phi_{a}+ 2\pi m_a,\delta\phi_{b}+ 2\pi m_b)$, which explicitly depend on the winding numbers $m_a$ and $m_b$. Then, the shortest total holding time reads
\begin{align}
    T_\mathrm{hold}(\delta\phi_{a},\delta\phi_{b})&= \min_{\substack{m_a,m_b\in \mathbb{Z}\\t_{A,B}\geq 0}}\Big[\nonumber\\&\hspace{0.85cm}t_{A}(\delta\phi_{a}+ 2\pi m_a,\delta\phi_{b}+ 2\pi m_b) \nonumber\\
    &\hspace{0.4cm}+ t_{B}(\delta\phi_{a}+ 2\pi m_a,\delta\phi_{b}+ 2\pi m_b)\Big],
\end{align}
where, to excluded non-physical negative holding times, we explicitly require
$t_{A,B}(\delta\phi_{a}+ 2\pi m_a,\delta\phi_{b}+ 2\pi m_b)\geq 0$.
Using $T_\mathrm{hold}(\delta\phi_{a},\delta\phi_{b})$, we compute the worst-case total hold-time for given laser-parameters as
\begin{equation}
    T_\mathrm{hold}^\mathrm{worst} = \max_{\delta\phi_{a,b}\in[0,2\pi]} T_\mathrm{hold}(\delta\phi_{a},\delta\phi_{b}).
\end{equation}
For a given set of laser parameters, the worst-case total gate-time to acquire all phase combinations is therefore given by
\begin{equation}
    T_\mathrm{gate}^\mathrm{worst}(\bm\Omega, \bm\Delta) = 2\left(T_\mathrm{ramps} + T_\mathrm{hold}^\mathrm{worst}\right),
\end{equation}
where $T_\mathrm{ramps} = T_1 +T_3+T_5 $.
This can then be used to find the optimal laser-parameters $(\bm\Omega^*, \bm\Delta^*)$ in order to get the minimal gate-duration to acquire all phases as
\begin{equation}
    (\bm\Omega^*, \bm\Delta^*) = \arg\min_{\substack{\Omega_k\in[0,\Omega_\mathrm{max}]\\\Delta_k\in[\Delta_\mathrm{min},\Delta_\mathrm{max}]}}  \lbrace T_\mathrm{gate}(\bm\Omega, \bm\Delta)\rbrace.
\end{equation}

\section{Time-optimal adiabatic ramps}
\label{appendix:QAB}

To obtain the individual time-optimal adiabatic ramps $(\Omega_0, \Delta_0)\!\!\rightarrow \!\!(\Omega_1, \Delta_1)$, we use a variational strategy, inspired by the quantum adiabatic brachistochrone method presented in Ref.~\cite{Rezakhani2009}. This section provides a simplified description of the algorithm. Further details will be presented in future work. 

To simplify the presentation, we only address the problem of adiabatically preparing a single target eigenstate of $H(\Omega_1, \Delta_1)$. However, the generalization to multiple target eigenstates is straightforward.
We consider an arbitrary continuous path $(f_\Omega(s), f_\Delta(s))$, parametrized by the dimensionless ``natural parameter’’ $0 \leq s \leq 1$, such that 

\begin{equation}
\begin{split}
  (f_\Omega(0), f_\Delta(0)) & = (\Omega_0, \Delta_0)\\
    (f_\Omega(1), f_\Delta(1)) & = (\Omega_1, \Delta_1)\;.
\end{split}
\end{equation}

For any total time $T$, this gives the time-dependent ramp 
\begin{equation}\label{eqn:t-ramp}
    \begin{split}
    \Omega_{f_\Omega, T}(t)&=f_\Omega(t/T)\\
    \Delta_{f_\Delta, T}(t)&=f_\Delta(t/T)
    \end{split}
    \;.
\end{equation}
We denote the fidelity between the target eigenstate $\ket{E_{\mathrm{targ}}(T)}$ and the final state $\ket{\psi(T)}$,  obtained using the ramp in Eq.~\eqref{eqn:t-ramp} to drive the system, as $\mathcal{F}_{T}[f_\Omega, f_\Delta]=|\braket{ \psi(T) | E_{\mathrm{targ}}(T) }|^2$.
The sole fidelity $\mathcal{F}_{T}[f_\Omega, f_\Delta]$ does not provide a measure of the adiabaticity of the ramp. Indeed, optimal quantum control (QOC)~\cite{Werschnik_JPhysB2007} and shortcuts to adiabaticity methods (STAs)~\cite{Guery-Odelin_RevModPhy2019} produce optimal drivings ($\mathcal{F}_{T}[f_\Omega, f_\Delta]=1$), which instead rely on a diabatic dynamics. To unequivocally quantify the adiabaticity of a path, we introduce the time functional
\begin{align}\label{eqn:time_functional}
    \mathcal{T}_{\epsilon}[f_\Omega, f_\Delta] &=\max\Big\{T: \mathcal{F}_{T}[f_\Omega, f_\Delta]\leq 1-\epsilon\Big\} \;.
\end{align}

Then, we say that the ramp in Eq.~\eqref{eqn:t-ramp} is $\epsilon$-adiabatic if $T\geq \mathcal{T}_{\epsilon}[f_\Omega, f_\Delta]$. 
By definition, an $\epsilon$-adiabatic ramp always results in a final fidelity $\mathcal{F}_{T}[f_\Omega, f_\Delta]\geq 1-\epsilon$. Moreover, this definition enforces an additional intuitive condition: slowing an adiabatic ramp does not deteriorate its adiabaticity. It thus excludes ramps which implement diabatic QOC or STAs.

The time functional $\mathcal{T}_{\epsilon}[f_\Omega, f_\Delta]$ also represents the minimum time necessary to realize an $\epsilon$-adiabatic ramp with the path $(f_\Omega(s), f_\Delta(s))$. In the absence of crossings in the spectrum of $H$, the adiabatic theorem~\cite{Messiah:book} assures that $\mathcal{T}_{\epsilon}[f_\Omega, f_\Delta]$ assumes a finite value. In particular using the results of Ref.~\cite{Jansen_JMatPhys2007}, we get the following rigorous upper bound:
\begin{align}\label{eqn:Time_functional_bound}
     \mathcal{T}_{\epsilon}[f_\Omega, f_\Delta] &\leq  
    \frac{1}{\epsilon}\bigg(\left.\frac{\lVert \partial_s H_{f_\Omega, f_\Delta}(s)\rVert}{g_{f_\Omega, f_\Delta}^2(s)}\right|_{s=0} \nonumber\\
    &\hspace{0.8cm}+ 
    \left.\frac{\lVert \partial_s H_{f_\Omega, f_\Delta}(s)\rVert}{g_{f_\Omega, f_\Delta}^2(s)}\right|_{s=1}\nonumber\\
    &\hspace{0.8cm} + \int_0^1\frac{\lVert \partial_s^2 H_{f_\Omega, f_\Delta}(s)\rVert}{g_{f_\Omega, f_\Delta}^2(s)}  \mathrm{d}s\nonumber\\
    &\hspace{0.8cm} + \int_0^1 \frac{\lVert \partial_s H_{f_\Omega, f_\Delta}(s)\rVert^2}{g_{f_\Omega, f_\Delta}^3(s)} \mathrm{d}s  \bigg)  
    \;,
\end{align}
where $ H_{f_\Omega, f_\Delta}(s) = H\left(f_\Omega(s), f_\Delta(s)\right)$ is the Hamiltonian path, and $g_{f_\Omega, f_\Delta}(s)$ is the energy gap that separates the target eigenstate form the rest of the spectrum. 

In principle, we can find the time-optimal $\epsilon$-adiabatic path minimizing the time functional $\mathcal{T}_{\epsilon}[f_\Omega, f_\Delta]$. However, due to the high computational cost of evaluating $\mathcal{T}_{\epsilon}[f_\Omega, f_\Delta]$, we must restrict the search to a smaller subset of paths. As suggested in Ref.~\cite{Roland_PRA2002,Rezakhani2009}, we consider the modified paths $(\tilde{f}^{(q)}_\Omega(s), \tilde{f}^{(q)}_\Delta(s))$, with 
\begin{align}
    \tilde{f}^{(q)}_{\Omega,\Delta}(s)&=f_{\Omega,\Delta}(\vartheta(s))\label{eqn:modified_path_1}\\ 
    \partial_s \vartheta (s)&=\frac{1}{A}\left(\frac{ g_{f_\Omega, f_\Delta}^2(s)}{\lVert\partial_s  H_{f_\Omega, f_\Delta}(s)\rVert}\right)^{q}\label{eqn:modified_path_2}\\
    A &= \int_0^1 \left(\frac{ g_{f_\Omega, f_\Delta}^2(s)}{\lVert\partial_s  H_{f_\Omega, f_\Delta}(s)\rVert}\right)^{q}\mathrm{d}s\label{eqn:modified_path_3}\;,
\end{align}
where $q$ is a real variational parameter.
Ref.~\cite{Jansen_JMatPhys2007} showed that, as long as $f_\Omega$ and $ f_\Delta$ satisfy some regularity conditions and $\frac{1}{
2}<q<1$, the modified path is such that $\mathcal{T}_{\epsilon}[\tilde{f}^{(q)}_\Omega, \tilde{f}^{(q)}_\Delta]\leq\frac{1}{\epsilon}\mathcal{O}\left( \frac{1}{\min_s [g_{f_\Omega, f_\Delta}(s)]}\right)$, which considerably improves the bound in Eq.\eqref{eqn:Time_functional_bound}. 

\begin{figure}[t]
\begin{centering}
\includegraphics[width=\columnwidth]{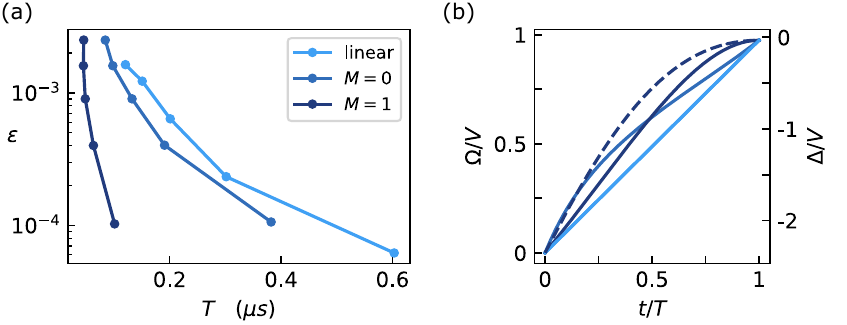}
\par\end{centering}
 \protect\caption{(a) Comparison of different adiabatic paths (linear, $M=0$, $M=1$) for the first ramp $(0, \Delta_\text{start})\!\!\rightarrow \!\!(\Omega_A, \Delta_A)$ in Fig.~3(a) of the main text. (b) Corresponding ramps for points with lowest infidelity $(\approx 10^{-4})$ in panel (a). Solid lines correspond to $\Omega(t)$ whereas dashed line corresponds to $\Delta(t)$. For cases where no dashed line is visible (linear, $M=0$) $\Omega$ and $\Delta$ curves fall on top of each other.}
\label{brachi_fig}
\end{figure}

We further restrict the search space by considering the subset of modified paths obtained interpolating $M$ points. Specifically, we consider the $M$ points $(\Omega_{s_m}, \Delta_{s_m})$ with $s_m=\frac{m}{M+1}$ and  $m=1,2,\dots,M$. Then, we use a third order spline interpolation to generate the paths $f_\Omega$, $f_\Delta$ such that $f_\Omega(s_m)=\Omega_{s_m}$, $f_\Delta(s_m)=\Delta_{s_m}$ for $m=0, 1,\dots,M,M+1$. Finally, we apply Eq.~\eqref{eqn:modified_path_1} to obtain the modified interpolated paths $\tilde{f}^{(q,\Omega_{s_1}, \dots, \Omega_{s_M})}_\Omega$, $\tilde{f}^{(q,\Delta_{s_1}, \dots, \Delta_{s_M})}_\Delta$, which now depend on $2 M+1$ variational parameters $(q,\Omega_{s_1}, \dots, \Omega_{s_M},\Delta_{s_1}, \dots, \Delta_{s_M})$.
Using the dual annealing algorithm~\cite{Tsallis_PhysA1996} (allowing for maximally 100 steps) , implemented in the SciPy library~\cite{pythonlib:SciPy_2020SciPy}, we numerically minimize the time functional of Eq.~\eqref{eqn:time_functional} over a subset of modified interpolated paths. The result is a variational time-optimal ramp, which we use as a building block for the two-pause-protocol gate described in the main text. 
Figure~\ref{brachi_fig}(a) shows how the choice of the adiabatic path affects the ramp infidelity $1-\mathcal{F}_T$ as a function of the ramp  duration. In particular, we compare ramps built using variational time-optimal paths (for $M=0,1$) and simple linear ramps
\begin{equation}\label{eqn:t-ramp_linear}
    \begin{split}
    \Omega^{\scriptstyle\mathrm{lin}}(t)&=\frac{t}{T}\Omega_0 + (1-\frac{t}{T})\Omega_1\\
    \Delta^{\scriptstyle\mathrm{lin}}(t)&=\frac{t}{T}\Delta_0 + (1-\frac{t}{T})\Delta_1
    \end{split}
    \;.
\end{equation}
Clearly, our method outperforms the simple linear ramp and improves significantly when increasing $M$. Figure~\ref{brachi_fig}(b) shows how the choice of path utilized the available parameter space. For the linear ramp the path in $(\Omega,\Delta)$ space is linear and the evolution speed is constant. In the $M=1$ case the $(\Omega,\Delta)$ path is still linear, however the evolution speed is adapted locally. For $M\geq1$ also the relation between $\Omega(t)$ and $\Delta(t)$ is not linear anymore, thus independently exploiting the available parameter space to speed up the ramp. 
We want to emphasize that utilizing variational time-optimal ramps considerably improves the performance of the Rydberg parity gate as described in the main text. The gate-performance results presented in the main text correspond to the choice $M=1$.

\section{Rydberg Potentials}
\label{appendix:Rydberg}

In this section we present calculations of the many body Rydberg potentials for the laser targeted $68S$ Rydberg states of ${}^{87}\mathrm{Rb}$, for up to four particles. Rydberg states $\ket{r} = \ket{n l j m_j}$ are characterized by their principal quantum number $n$, orbital angular momentum $l$, total angular momentum $j$, and projection, $m_j$, of $j$ onto the quantization axis. These states $\ket{r}$ are the eigenstates of the valence electron of an alkali atom, described by a single-particle Hamiltonian $\hat{H}_A$ which includes e.g. the interaction with the atomic core, as well as external fields. 

\begin{figure}[h]
\begin{centering}
\includegraphics[width=\columnwidth]{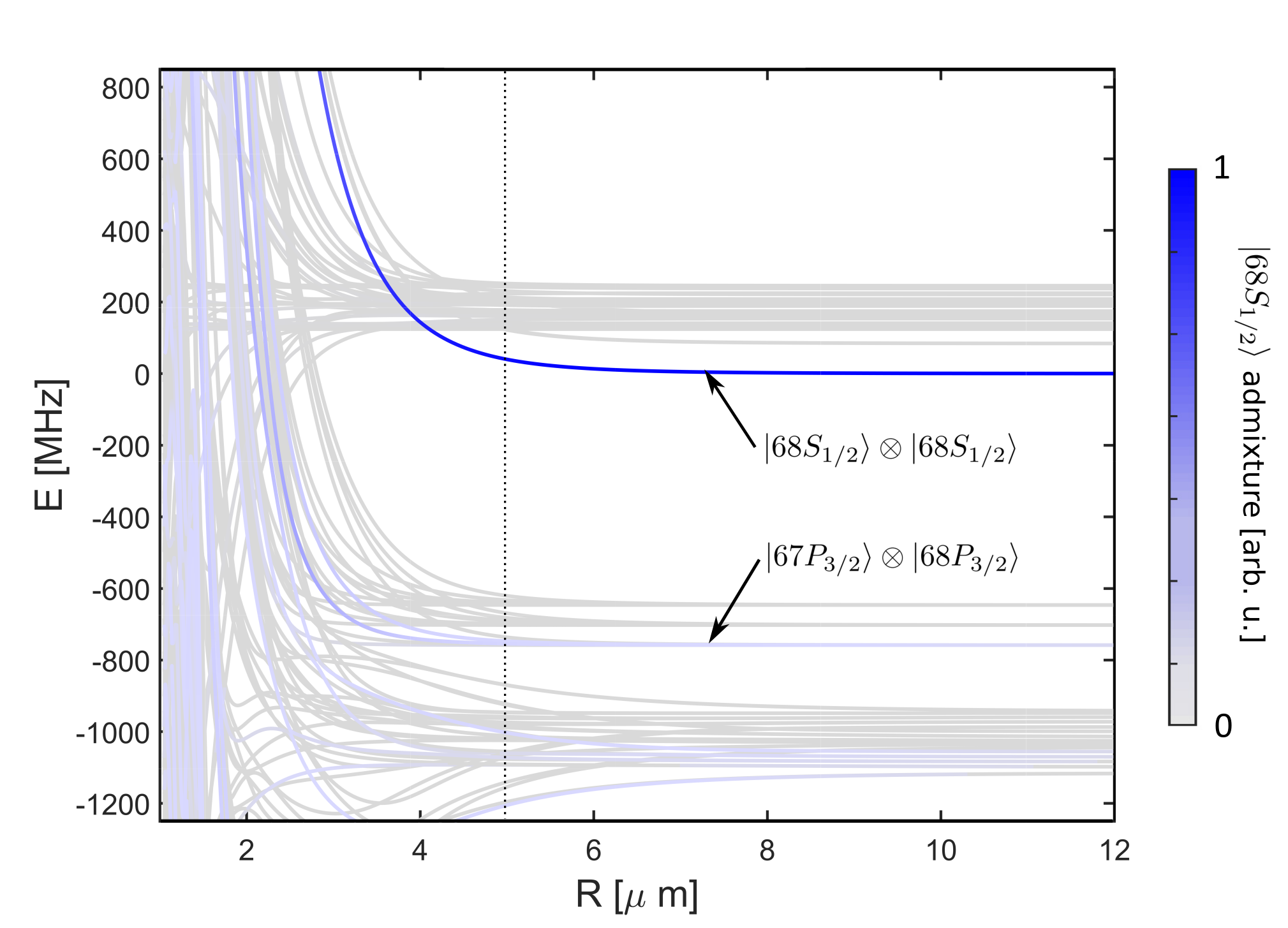}
\par\end{centering}
 \protect\caption{Two particle eigenstates around the $\ket{68S}\otimes\ket{68S}$ Rydberg states, as a function of interparticle separation $R$. For each eigenstate $\ket{\psi}$, blue coloring indicates admixture $\braket{\psi_0|\psi}$, with $\ket{\psi_0} = \ket{68S_{1/2}, m_j = -1/2}^{\otimes 2}$. This admixture is indicative of the strength of the laser coupling from the ground state.}
\label{potentials_fig}
\end{figure}

When we consider multiple particles $i = 1,\dots,4$, at positions $\mathbf{r}_i$, and with interparticle separation vectors $\mathbf{R}_{ij} = \mathbf{r}_i - \mathbf{r}_j$, there are additional dipole-dipole couplings of the form
\begin{equation}
\hat{V}_{dd}^{(i, j)} = \frac{\bm{\mu}_i \cdot \bm{\mu}_j}{R_{ij}^3} - \frac{3 (\bm{\mu}_i \cdot \mathbf{R}_{ij})(\bm{\mu}_j\cdot \mathbf{R}_{ij})}{R_{ij}^5},
\end{equation}
where $\bm{\mu}_i$ is the dipole transition operator for atom $i$. In the present paper we only consider interatomic distances where higher order multipole couplings (quadrupole, octupole, etc.) can be ignored.

The Hamiltonian describing the Rydberg states of a $2 \times 2$ plaquette then becomes
\begin{equation}\label{eq:Hryd}
\hat{H}^{(\mathrm{Ryd})} = \sum_i \hat{H}_A^{(i)} + \sum_{i<j}\hat{V}_{dd}^{(i, j)},
\end{equation}
where $\hat{H}_A^{(i)}$ is the single particle Hamiltonian, acting on particle $i$. After selecting a suitable set of basis states $\ket{n l j m_j} \otimes \ket{n'l'j'm_j'} \otimes \dots$, we can diagonalize the Hamiltonian (\ref{eq:Hryd}) to obtain the many-body spectrum. 

Figure \ref{potentials_fig} shows the eigenstates for two particles, diagonalized around the $\ket{68S}\otimes\ket{68S}$ states. The particles are oriented on the $z$-axis, in the presence of a magnetic field of $B = 21\,\mathrm{Gauss}$, oriented along the same axis. Since the Hamiltonian (\ref{eq:Hryd}) couples pair states $\ket{n l j m_j} \otimes \ket{n'l'j'm_j'}$, the eigenstates become superpositions of these basis states. Blue coloring indicates the admixture of the laser-targeted $\ket{68S} \otimes \ket{68S}$ states ($n = 68, l = 0, j = 1/2, m_j = -1/2$). The strong, blue-colored line shows a clear $1/R^6$ distance dependence of the van der Waals interactions between the $68S$ states. At a distance of $5\,\mathrm{\mu m}$, the interaction energy $E = 40 \,\mathrm{MHz}$, the number used in the main text. At an energy of $-750\,\mathrm{MHz}$, another band of states is visible, corresponding to eigenstates predominantly composed of $P$ states, with a small ($\sim 1\%$) admixture of $68S$ states. 

\begin{figure}[b]
\begin{centering}
\includegraphics[width=\columnwidth]{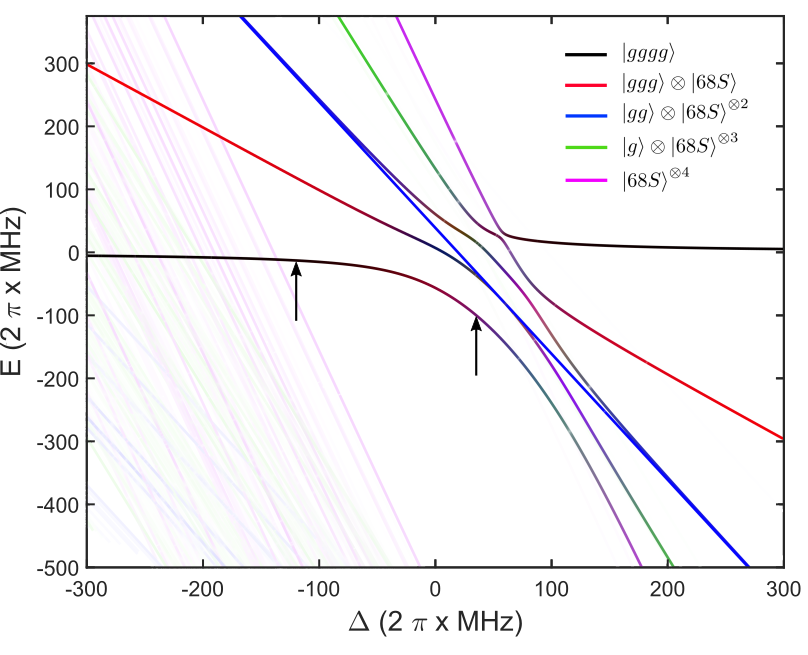}
\par\end{centering}
 \protect\caption{Many-body eigenstates of a $2 \times 2$ plaquette, in the presence of a laser coupling between ground states $\ket{g}$ and the Rydberg manifold, with Rabi frequency $\Omega = 2\pi \times 40\, \mathrm{MHz}$, and plotted as a function of the laser detuning $\Delta$. Coloring indicates the number of excitations, with color intensity representative for the relative amount of $68S$-state admixture, as compared to the total Rydberg fraction, and with white being zero. Vertical arrows indicate the gate operation range.}
\label{pgl_fig}
\end{figure}

For a proper gate operation, we need to ensure that there are no states (like those at $-750\,\mathrm{MHz}$) other than the laser-targeted $68S$ states that become resonant for detunings within the gate operation regime. To this end, we extend the state space of single particle states $\ket{n l j m_j}$ with a noninteracting ground state $\ket{g}$, and extend the Hamiltonian (\ref{eq:Hryd}) with the laser coupling operator
\begin{equation}
\hat{V}_L = \sum_i \sum_{n,l,j,m_j} \frac{\Omega_{n,l,j,m_j}}{2} \ket{n, l, j, m_j}\bra{g} + \mathrm{H.c.},
\end{equation}
where $\Omega_{n,l,j,m_j}$ is the effective Rabi frequency between the ground state $\ket{g}$ and Rydberg state $\ket{n l  j m_j}$. Due to selection rules, this Rabi frequency is mostly zero, but it is nonzero not only for the $68S$ states, but e.g. also for other $S$ states. For simplicity, we assume that $\Omega_{n,l,j,m_j}$ is either zero, or a fixed value $\Omega$ when not forbidden by selection rules.

We now consider $4$ particles forming a $2\times2$ plaquette in the tetrahedron configuration as discussed in the main text, with particles positioned at $5\,\mathrm{\mu m}$ distance, and in the presence of a magnetic field of $B = 21\,\mathrm{Gauss}$. We subsequently diagonalize the Hamiltonian $\hat{H}^{(\mathrm{Ryd})} + \hat{V}_L$ for a range of detunings, and $\Omega = 2\pi \times 40\, \mathrm{MHz}$. Figure~\ref{pgl_fig} shows the resulting many-body spectrum. This figure is a generalization of Fig.~\ref{fig2new} in the main text, where only a single possible Rydberg level was considered. In Fig.~\ref{pgl_fig}, coloring indicates the number of Rydberg excitations in each state, with color intensity proportional to the admixture of laser-excitable $68S$ states, with white indicating zero overlap. Besides the expected structure cf. Fig.~\ref{fig2new} of the main text, we see additional eigenstates appearing. In particular, a faintly visible manifold of double, triply and quadruply excited states appears at negative detunings. These states are products of the states with $\ket{67P}\otimes\ket{68P}$ character (as discussed above) with $\ket{g}$ and $\ket{68S}$ states. 
Arrows indicate the eigenstate and detuning range that our gate example of the main text operates on. Clearly, any resonances with the spurious states are avoided. For smaller detuning $\Delta < 2\pi \times -135\, \mathrm{MHz}$, small resonances with the targeted eigenstate occur, although they are very narrow (not resolved in the figure, $< 1\, \mathrm{kHz}$).

\section{Open system dynamics and average gate fidelity}
\label{appendix:ME}
In the following we describe the model used to analyze the parity gate subject to radiative decay of population in the Rydberg manifold [see Fig.~3(b)]. Our model includes spontaneous decay from the Rydberg state $\ket{r}$ with decay rate $\gamma_r$ to the five hyperfine ground state levels of ${}^{87}\mathrm{Rb}$, where an uncoupled state $\ket{d}$ represents the three hyperfine ground states outside the qubit basis~\cite{Saffman2020}. This gives rise to decay branching ratios of $b_{r\uparrow} = b_{r\downarrow}=1/5$ and $b_{rd} = 3/5$. Note, that population leakage into the state $\ket{d}$ is treated as an uncorrectable error. The dynamics w.r.t.\@ the original state-space (with basis states $\lbrace\ket{\downarrow},\ket{\uparrow},\ket{r}\rbrace^{\otimes 4}$) can be described by the master equation 
\begin{equation}\label{eq:ME}
    \frac{\mathrm{d}\rho}{\mathrm{d}t} = -i[H,\rho] + \mathcal{D}[\rho],
\end{equation}
Where $H$ is the single-plaquette Hamiltonian described in the main text and the dissipative term $\mathcal{D}[\rho]$ is given by
\begin{eqnarray}
    \mathcal{D}[\rho] = \sum_{k = 1}^4\sum_{l= \uparrow,\downarrow}L_l^{(k)}\rho L_l^{\dagger (k)} - \frac{1}{2}\lbrace D_r,\rho\rbrace,
\end{eqnarray}
with the decay operators $L_{l}^{(k)} =  \sqrt{b_{rl}\gamma_r}\ket{l_k}\bra{r_k}$,  $D_r =\sum_{k=1}^4\gamma_r\ket{r_k}\bra{r_k}$~\cite{Saffman2020}. In order to reduce the computational complexity of solving Eq.~\eqref{eq:ME}, we analyze the gate performance within the permutation symmetric superoperator subspace. Denoting the Liouville superoperator on the symmetric subspace as $\tilde{\mathcal{L}}$ and $\tilde{\mathcal{U}}_{\squaredots}$ as the symmetrized superoperator representation of the ideal gate operation given by $U_{\squaredots}^\dagger\rho U_{\squaredots}$, we calculate the average gate fidelity as 
\begin{equation}
    \overline{\mathcal{F}} = \frac{d + \mathrm{tr}\!\left(\tilde{\mathcal{U}}_{\squaredots}^\dagger \tilde{\mathcal{L}}\right)}{d(d+1)},
\end{equation}
where $d$ is the dimension of the quantum channel~\cite{Nielsen2002, Wood2015}.

\bibliography{main.bbl}

\end{document}